# Wafer-scale Semiconductor Grafting: Enabling High-Performance, Lattice-Mismatched Heterojunctions

Jie Zhou[1], Qiming Zhang[1], Jiarui Gong[1], Yi Lu[1], Yang Liu[1], Haris Abbasi[1], Haining Qiu[1], Jisoo Kim[1], Wei Lin[1], Donghyeok Kim[1], Yiran Li[1], Tien Khee Ng[2], Hokyung Jang[1], Dong Liu[1], Haiyan Wang[3], Boon S. Ooi[2,a)], and Zhenqiang Ma[1,a)]

*[1]Department of Electrical and Computer Engineering, University of Wisconsin-Madison, Madison, Wisconsin, 53706, United States*

*[2]Department of Electrical and Computer Engineering, King Abdullah University of Science and Technology, Thuwal 23955-6900, Saudi Arabia*

*[3]School of Materials Engineering, Purdue University, West Lafayette, Indiana 47907, United States*

[a)] Author to whom correspondence should be addressed. Electronic mail: mazq@engr.wisc.edu, boon.ooi@kaust.edu.sa

# Abstract

Semiconductor heterojunctions are foundational to many advanced electronic and optoelectronic devices. However, achieving high-quality, lattice-mismatched interfaces remains challenging, limiting both scalability and device performance. Semiconductor grafting offers a promising solution by directly forming electrically active, lattice-mismatched heterojunctions between dissimilar materials. However, its scalability and uniformity at the wafer level have yet to be demonstrated. This work demonstrates the achievement of highly uniform, reproducible results across silicon, sapphire, and gallium nitride (GaN) substrates using wafer-scale semiconductor grafting. To illustrate this scalability, we conducted an in-depth study of a grafted Si/GaN heterojunction, examining band alignment through X-ray photoelectron spectroscopy and confirming crystallinity and interfacial integrity with scanning transmission electron microscopy. The resulting *p-n* diodes exhibit significantly enhanced electrical performance and wafer-scale uniformity compared to conventional approaches. This work establishes wafer-scale semiconductor grafting as a versatile and scalable technology, bridging the gap between laboratory-scale research and industrial manufacturing for heterogeneous semiconductor integration, and paving the way for novel, high-performance electronic and optoelectronic devices.

**Introduction**

Semiconductor heterojunctions are among the most transformative inventions of the last century, underpinning a range of devices like lasers, LEDs, and transistors that have profoundly impacted society. Herbert Kroemer's foundational work established the theoretical basis for enhancing electronic and optoelectronic device performance by leveraging heterojunctions—interfaces between dissimilar semiconductors [1–3]. These "Kroemer heterojunctions" provide indispensable measures to engineer device properties by creating an energy band discontinuity at the junction, which can confine or accelerate charge carriers. However, forming high-quality heterojunctions typically requires lattice matching to minimize interfacial defects, a constraint that has restricted heterojunction research primarily to a limited set of compatible materials, including Si, GaAs, InP, and nitrides. Many of these applications have already transitioned from research laboratories to practical implementations, making a tangible societal impact [4–7].

While lattice-matched heterojunctions have yielded remarkable successes, lattice-mismatched heterojunctions hold untapped potential for device innovation. Attempts to create high-performance mismatched heterojunctions through techniques such as heteroepitaxy and direct bonding face fundamental challenges: lattice mismatch inherently introduces interfacial defects that degrade charge transport efficiency, limiting device performance.

Semiconductor grafting [8] is an emerging technology that addresses these limitations. Semiconductor grafting enables creation of heterojunctions between arbitrary semiconductors, circumventing the limitations of lattice matching and offering electrically active interfaces, which in turn shades light on new designing freedom for heterogeneous semiconductor devices. Leveraging this technique, we have demonstrated various grafted heterojunction devices, including electronic diodes [9–13], transistors [14,15], optoelectronic LEDs [11,16–18], and lasers [19], exhibiting superior performance comparable to or surpassing their lattice-matched counterparts that are conventionally realized by heteroepitaxy or direct bonding.

Thus far, semiconductor grafting has demonstrated success in millimeter-scale chips, exhibiting repeatable and reliable device process flows and optimal performance. However, akin to the development of mature lattice-matched epitaxy and heteroepitaxy, large-area fabrication with wafer-scale device uniformity must be achieved before this emerging technology can advance towards mass production and commercialization. Crucially, achieving scalability in semiconductor grafting will not only enable cost-effective production but may also improve device performance through uniform, one-step grafting. This may further facilitate the transition from single-device studies to on-chip device interconnection and integration research.

Typically, semiconductor grafting utilizes transfer printing techniques involving single-crystalline semiconductor nanomembranes (NMs) to establish abrupt heterojunctions between the grafted NMs and the hosting semiconductors. Although some existing methods have been proposed to facilitate large-scale production and transfer printing of NMs or thin films [20–24], semiconductor grafting, which necessitates an atomically clean interface for optimal device performance, poses more stringent requirements. Achieving wafer-scale grafting requires overcoming substantial challenges related to the structural integrity and precise alignment of NMs at large scales. As NMs increase in size, they become more susceptible to wrinkling, stress-induced deformation, and alignment errors, making standard transfer techniques [8] insufficient. Specialized methods are needed to preserve NM stability and ensure precise placement, as even minor misalignments or structural imperfections can significantly impair device functionality and performance. Additionally, achieving uniformity in the high-fidelity membrane detachment process and maintaining surface cleanliness across the wafer are equally essential; any variations in these steps can introduce defects, compromising the quality and reliability of the heterojunctions. Addressing these challenges represents a major leap from previous millimeter-scale grafting to the demanding realm of wafer-scale grafting, enabling the high-quality interfaces necessary for high-performance grafted devices.

To our knowledge, no previous studies have reported grafting or demonstrated device uniformity at the wafer scale. In this work, we present wafer-scale grafting technology for the first time. Si NMs up to two inches in size were successfully grafted onto various substrates, including silicon, sapphire, and GaN. Using the grafted Si/GaN heterojunction—characterized by distinct lattice structures (diamond versus wurtzite) and a significant lattice mismatch of 70.2%—we fabricated and characterized wafer-scale *p-n* diodes. The devices exhibited excellent and consistent diode performance across the entire wafer, confirming the feasibility and reliability of the semiconductor grafting process. This breakthrough opens new opportunities for wafer-scale heterogeneous electronics and photonics, advancing the development and commercialization of these innovative semiconductor devices.

## Results and Discussion

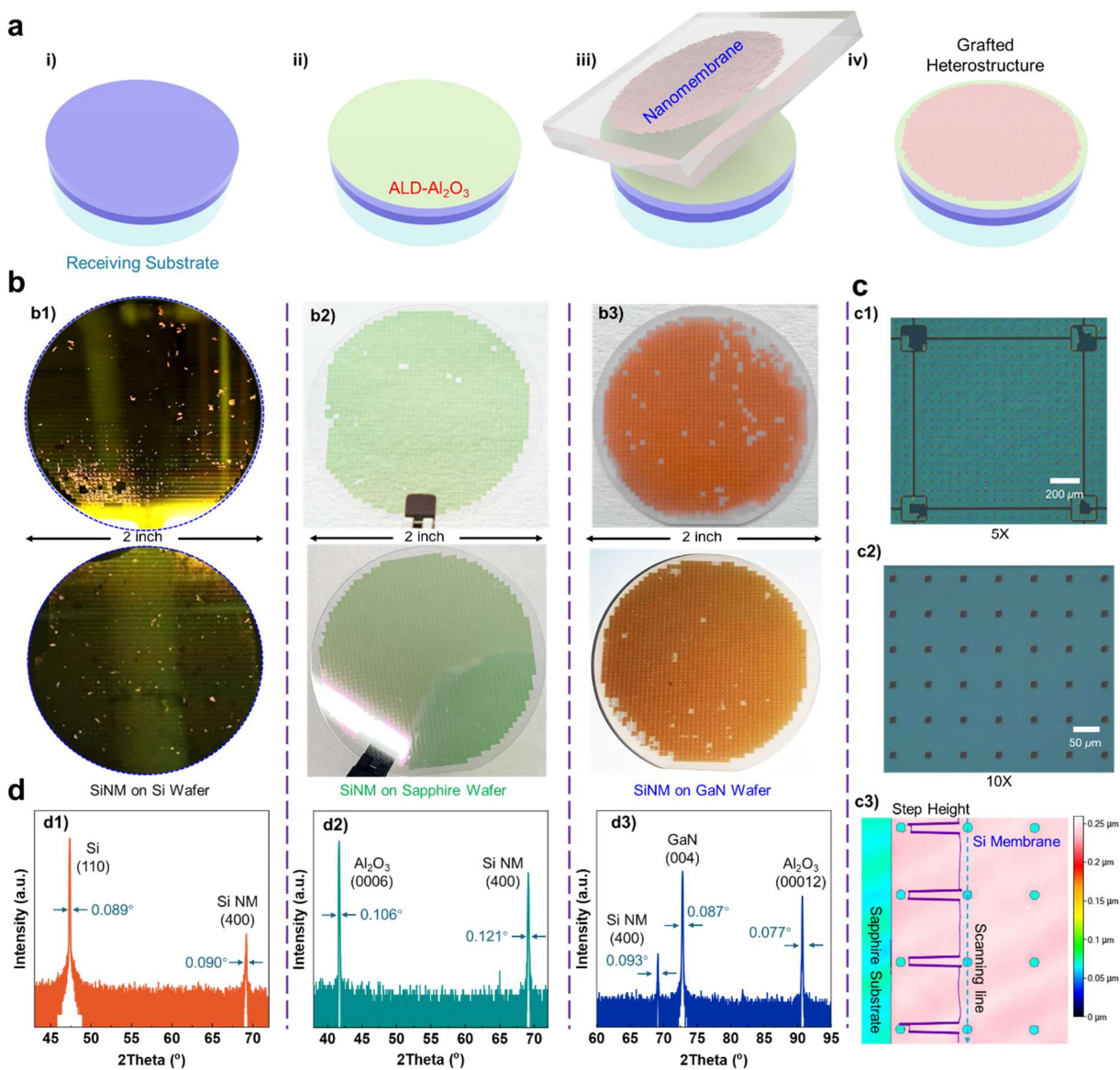

**Fig. 1 Conceptual illustration, real-world images, and characterizations of wafer-scale grafted semiconductor heterojunctions.** (a) Schematic of the semiconductor grafting process: i) and ii) surface passivation of the receiving crystalline substrate with an ultrathin oxide layer, iii) transfer of freestanding NMs, and iv) chemical bonding to form the heterostructure. (b) Macroscopic images of two-inch wafers showing Si NMs grafted onto Si, sapphire, and GaN substrates, highlighting the high transfer yield and reproducibility of the grafting process. (c) Microscopic image of a grafted Si NM on a sapphire substrate, demonstrating its ultra-thin, conformal, and smooth properties. (d) X-ray diffraction (XRD) analysis of the grafted

heterojunctions on different substrates, confirming the single crystalline nature of both the grafted layers and underlying substrates.

The conceptual representation of semiconductor grafting technology is illustrated in **Fig. 1(a)**. The process begins with the preparation of the receiving substrate ('Semiconductor A') as shown in **Fig. 1(a i)**. To avoid direct contact between the dissimilar crystal lattices of the two semiconductors, the receiving substrate is typically coated with an ultrathin oxide (UO) layer via atomic layer deposition (ALD), as depicted in **Fig. 1(a ii)**. This amorphous UO layer serves two critical functions: it acts as both an interface passivation layer and a quantum tunneling layer [8]. In the next step of the wafer-scale grafting process, a large, wafer-scale, freestanding single-crystalline semiconductor NM, referred to as 'Semiconductor B,' is carefully released and transfer-printed onto the ALD-passivated substrate, as shown in **Fig. 1(a iii)**. Handling and aligning such a large, delicate NM presents significant challenges, as it is highly susceptible to wrinkling, deformation, and misalignment. To overcome these challenges and achieve high-quality wafer-scale grafting, a combination of technical innovations is employed to preserve NM integrity and alignment. First, a pixelized NM design with anchors secures each individual NM unit during the release process, preventing unintended displacements or deformation that could arise from internal stress release during the undercutting process. Each NM unit, measuring 1 mm × 1 mm, is surrounded by four anchors that ensure stability and enable a controlled detachment while reducing the risk of wrinkles and delamination during pick-up. To further enhance precision and uniformity, a specialized honeycomb-patterned PDMS stamp is employed, which provides controlled adhesion and minimizes mechanical stress on the NM during transfer. This honeycomb structure effectively mitigates the risks of wrinkling and displacement that typically occur with conventional flat, non-textured PDMS stamps, particularly at larger scales. Additionally, the honeycomb pattern allows for air escape during the transfer, reducing the likelihood of trapped air pockets that could compromise NM bonding quality. Once the NM is precisely aligned with the receiving substrate, a gentle but firm application of pressure secures it onto the ALD-coated surface, creating an initial van der Waals bond that holds the NM in place without compromising its crystalline quality. This meticulous approach to NM handling and transfer is essential to achieving wafer-scale grafting uniformity and ensuring high-quality interfaces. See **Supplementary Information** for detailed information containing structure illustration and microscope images of honeycomb PDMS, and pixelization design of Si NMs.

The grafting process is completed with thermal annealing, which transforms the initial van der Waals bond into a strong chemical bond, as demonstrated in **Fig. 1(a iv)**.

To showcase the practical implementation of this technology, we present the successful wafer-scale grafting of Si NMs onto diverse substrates, including Si, sapphire, and GaN, as illustrated in **Figs. 1(b1)** to **1(b3)**, respectively. This grafting process, which builds upon the fundamental principles of semiconductor grafting, has been successfully reproduced with high throughput, achieving a grafting yield consistently exceeding 90% across all six wafers. This high yield underscores the reliability and repeatability of our wafer-scale grafting method. See the **Experimental Section** and **Supplementary Information** for a detailed description of the wafer-scale grafting process. **Figure 1(c1)** and **Fig. 1(c2)** show differential interference contrast (DIC) microscope images of a grafted Si NM unit at 5X and 10X magnifications, respectively. Optical profiling of a grafted Si NM was performed using a Profilm3D® profilometer. This profiling included a scanning line across the mesh holes of the Si NM, with corresponding step-depth measurements. The grafted NMs are characterized to be only ~180 nm thin. As shown in the microscopic and optical profiling images (**Figs. 1(c1)** to **1(c3)**), the Si NMs remain seamlessly bonded to the substrates after wafer-scale grafting and thermal bonding, despite the significant mismatch in both lattice constants and thermal expansion coefficients with the receiving substrate. This seamless bonding indicates high fidelity, completeness, and conformity of the grafted Si NMs, with no observable localized defects.

To confirm the preservation of single crystallinity after grafting, X-ray diffraction (XRD) was used to analyze three types of grafted wafers: Si NM(400)/Si substate(110), Si NM(400)/sapphire ($Al_2O_3$ 0006), and Si NM(400)/GaN(004), and **Fig. 1(d)** presents the 2 theta-omega XRD spectra for each type of wafer. The full width at half maximum (FWHM) values for each wafer type are marked in **Figs. 1(d1)** to **1(d3)**, respectively. These exceptionally narrow FWHM measurements confirm the single crystallinity of both the grafted Si NMs and the substrates.

These characterizations unequivocally confirm the exceptional quality, uniformity, and scalability of the wafer-scale semiconductor grafting process, thus offering a novel approach to reliably integrate Si NMs onto diverse substrates, ensuring high crystallinity and defect-free bonding.

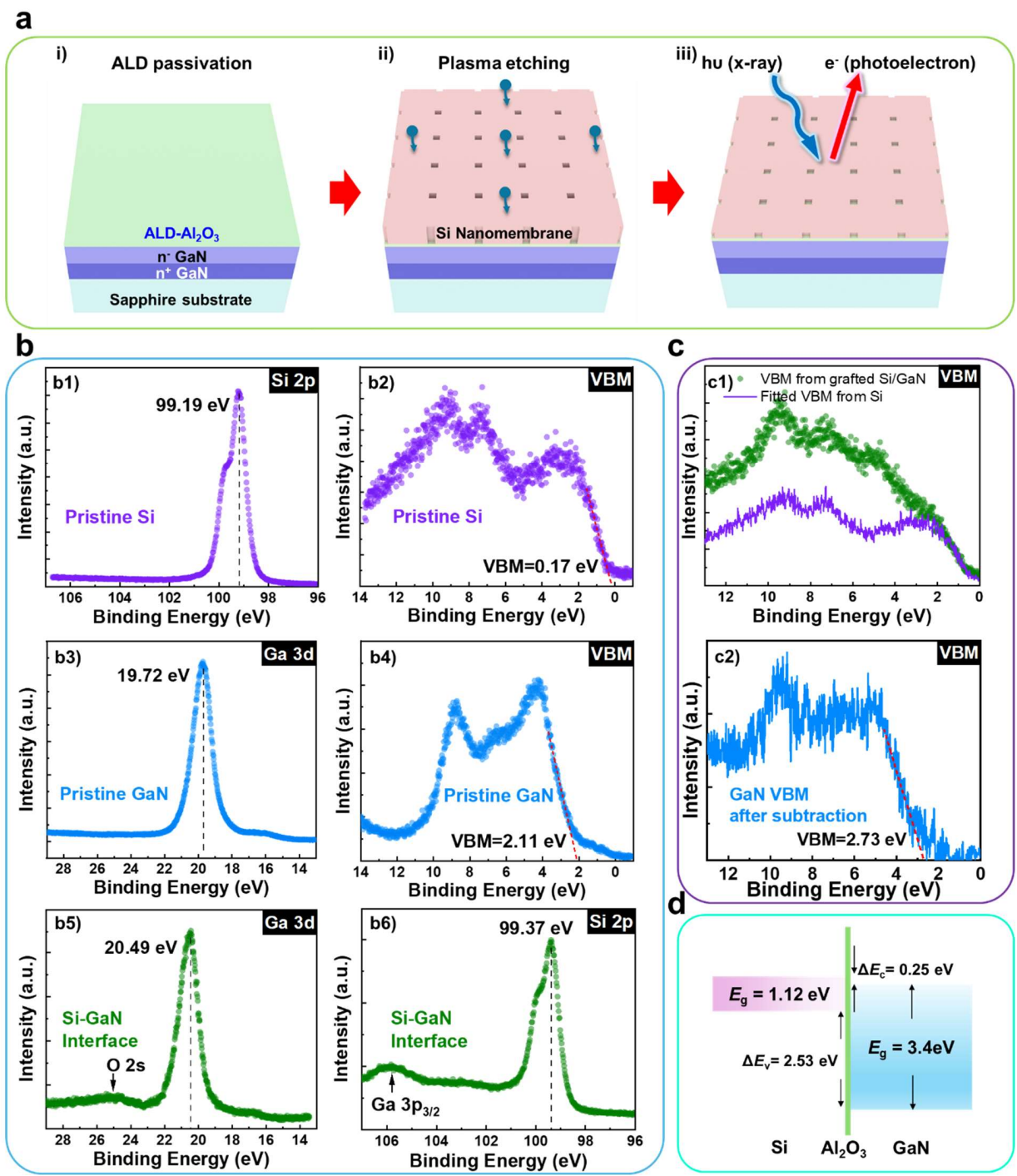

**Fig. 2 X-ray photoelectron spectroscopy (XPS) analysis of band alignment within the grafted Si/GaN heterojunction.** Scatters represent raw XPS data, while lines show the fitted or processed results. (a) Schematic of the XPS sample fabrication process: i) An n-GaN substrate is coated with ultrathin ALD-Al₂O₃, ii) a 180 nm Si NM is grafted onto the substrate, and iii) the Si NM is thinned down to ~10 nm. (b) Band alignment analysis using core level method. b1) Si 2*p*

XPS spectrum obtained from the SOI sample with extraction of the Si 2p peak position. b2) Valence band XPS spectrum of the SOI sample with extraction of the VBM. b3) Ga 3d XPS spectrum collected from the pristine GaN substrate with extraction of the Ga 3d peak position. b4) Valence band XPS spectrum of the pristine GaN substrate with extraction of the valence band maximum (VBM). b5) Ga 3d XPS spectrum of the grafted Si/GaN heterojunction sample with extraction of the Ga 3d peak position. b6) Si 2p XPS spectrum of the Si/GaN heterojunction sample with extraction of the Si 2p peak position. (c) Band alignment construction using valence band spectrum analysis. c1) Valence band XPS spectrum of the grafted Si/GaN sample (green scatters), with the purple curve representing the fitted spectrum contribution from the Si valence band b2). c2) The valence band XPS spectrum of GaN obtained from subtracting the contribution of Si from the grafted Si/GaN heterojunction c1), with the extraction of the VBM value (2.73 eV). (d) Band diagram constructed for the grafted monocrystalline Si/GaN heterojunction from the XPS measurement.

Additionally, we characterized the band alignment of the grafted Si/GaN heterojunction using XPS, following previously reported methods [9,25] with the results detailed in **Fig. 2**. To ensure electron escape from the grafted heterointerface for detection, we reduced the thickness of the Si NMs to around 10 nm, as illustrated in **Fig. 2(a)**. See the **Experimental Section** for further procedural details.

We first constructed the band alignment using the core level method [25,26], with the results summarized in **Fig. 2(b)**. XPS spectra of the Si 2p core level and valence band were obtained from the pristine SOI sample (**Figs. 2(b1)** and **2(b2)**), while those for the Ga 3d core level and valence band were measured from the pristine GaN substrate (**Figs. 2(b3)** and **2(b4)**). From the grafted Si/GaN heterojunction, we measured the Ga 3d and Si 2p core levels (**Figs. 2(b5)** and **2(b6)**). Using these data and reported bandgap values for Si and GaN, we calculated the band alignment following the method by Kraut et al. [26].

To cross-verify our measurements and calculations, we also employed the valence band method [9,25,27]. As shown in **Fig. 2(c1)**, the valence band XPS spectrum of the grafted Si/GaN heterojunction (green scatters) contains contributions from both Si (**Fig. 2(b2)**) and GaN (**Fig. 2(b4)**); the Si contribution (purple curve) was fitted and subtracted to isolate the GaN signal, allowing extraction of the VBM value (**Fig. 2(c2)**). The valence band offset between Si and GaN was determined from the binding energy difference between their respective VBM values (**Figs.**

**2(b2)** and **2(c2)**). **Figure 2(d)** summarizes the band alignment results from both methods, which are in excellent agreement, with a measurement error margin of approximately ± 0.1 eV. See **Supplementary Information** for the band diagram parameter calculations of these two methods.

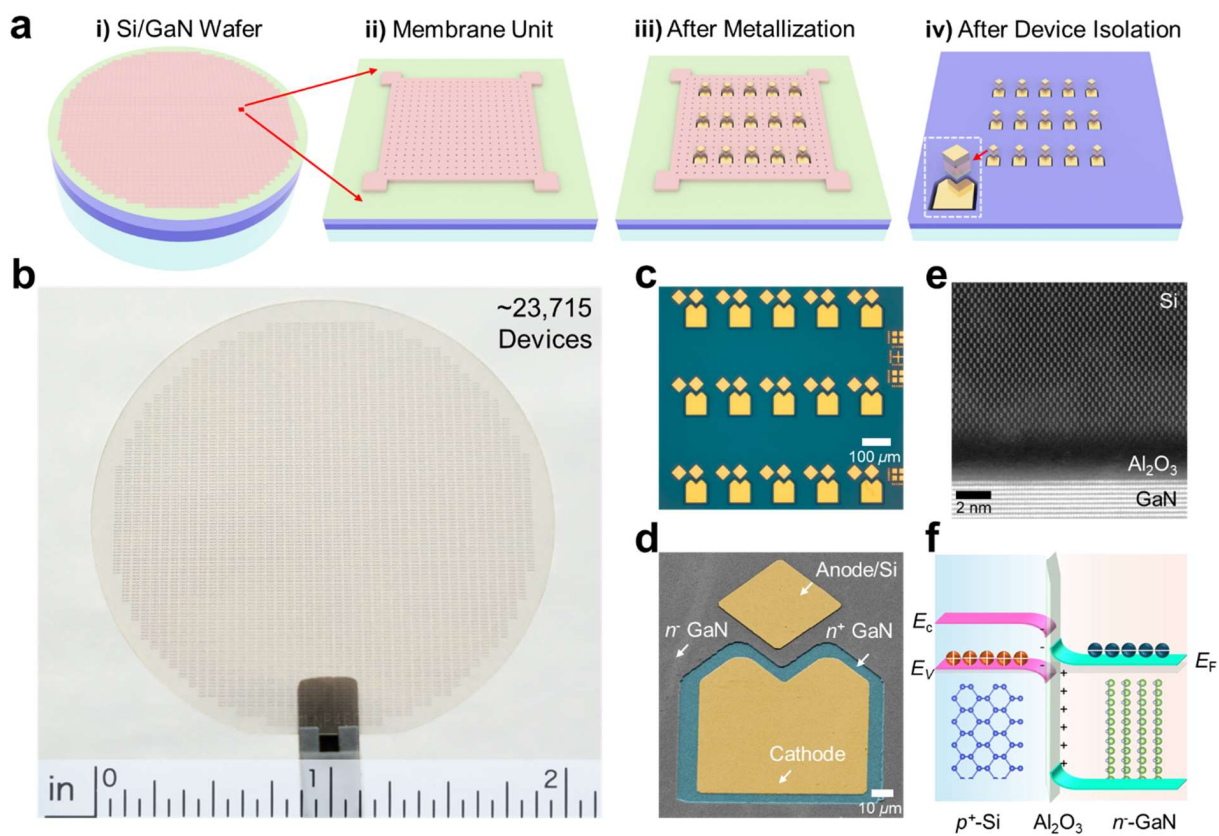

**Fig. 3 Macroscopic and microscopic images of the fully fabricated two-inch Si/GaN wafer and *p-n* diodes.** (a) Schematic illustration of the fabrication process for Si/GaN heterojunction diodes on a two-inch wafer. (b) Photograph of the completed Si/GaN wafer, containing approximately 23,715 devices. (c) Optical micrograph of a single unit, consisting of nine Si/GaN diodes. (d) Pseudo-colored scanning electron microscope (SEM) image of a single diode. (e) High-angle annular dark-field scanning transmission electron microscope (HAADF-STEM) image of the Si/GaN heterointerface, with the amorphous ALD-$Al_2O_3$ interlayer appearing as a darkened region. (f) Schematic of the crystallinity and band alignment for the grafted Si/GaN diodes.

To evaluate device performance and uniformity, we performed wafer-scale fabrication and characterization of lattice-mismatched Si/GaN *p-n* diodes, with a simplified overview of the fabrication process shown in **Fig 3(a)**. This process includes metallization of both the grafted $p^+$

Si NMs and the $n^+$ GaN epilayer, and device isolation by dry etching to remove excess Si NMs. See the **Experimental Section** for full details of the wafer-scale device fabrication process, and the **Supplementary Information** for structural schematics and doping profile details of the fabricated Si/GaN diodes.

Beyond the fabrication steps, careful management of the thermal budget during both grafting and diode fabrication is crucial for achieving optimal device performance. A well-controlled thermal budget is essential for preserving a high-quality interface by maintaining the functionality of the ultrathin oxide (ALD-Al$_2$O$_3$) interlayer, accommodating the thermal expansion mismatch between the grafted layer and the substrate, and ensuring reliable ohmic contact formation, ultimately enhancing performance of grafted devices. Variations in the thermal budget can significantly affect the density of interfacial traps ($D_{it}$) and key diode characteristics, such as the ideality factor and leakage current. To investigate this impact, we systematically analyzed the performance of grafted diodes across an individual Si/GaN wafer at different fabrication stages, with detailed performance variations and statistical trends provided in the **Supplementary Information**.

**Figure 3(b)** shows an overview of the fully fabricated wafer with its dimensions indicated. **Figure 3(c)** presents a magnified view of a single device section, captured using a microscope, while **Figure 3(d)** shows a scanning electron microscope (SEM) image highlighting the double-layered device configuration, where the anode is positioned on the $p^+$ Si NM and the cathode on the exposed $n^+$ GaN layer. High-resolution HAADF-STEM images of the Si/GaN heterojunction interface (**Fig. 3(e)**) reveal clean, sharp interfaces with the amorphous Al$_2$O$_3$ layer appearing as a dark region. See **Supplementary Information** for additional low-magnification images and EDX elemental scans, which confirm the lateral uniformity of the entire grafted junction. **Figure 3(f)** schematically represents the crystalline orientations and structures—diamond lattice for Si and wurtzite lattice for GaN—observed in the STEM micrographs, along with the electronic band alignment of the grafted Si/GaN $p$-$n$ junction at the thermal equilibrium state.

**Figure 3** demonstrates the successful achievement of high-quality, high-yield Si NM grafting at both the wafer and junction levels, with atomically clean and abrupt heterointerfaces. Notably, we observed no significant device loss across the wafer-scale fabrication process, except for minor yield reductions in some peripheral regions attributed to unavoidable misalignment (see **Supplementary Information** for details).

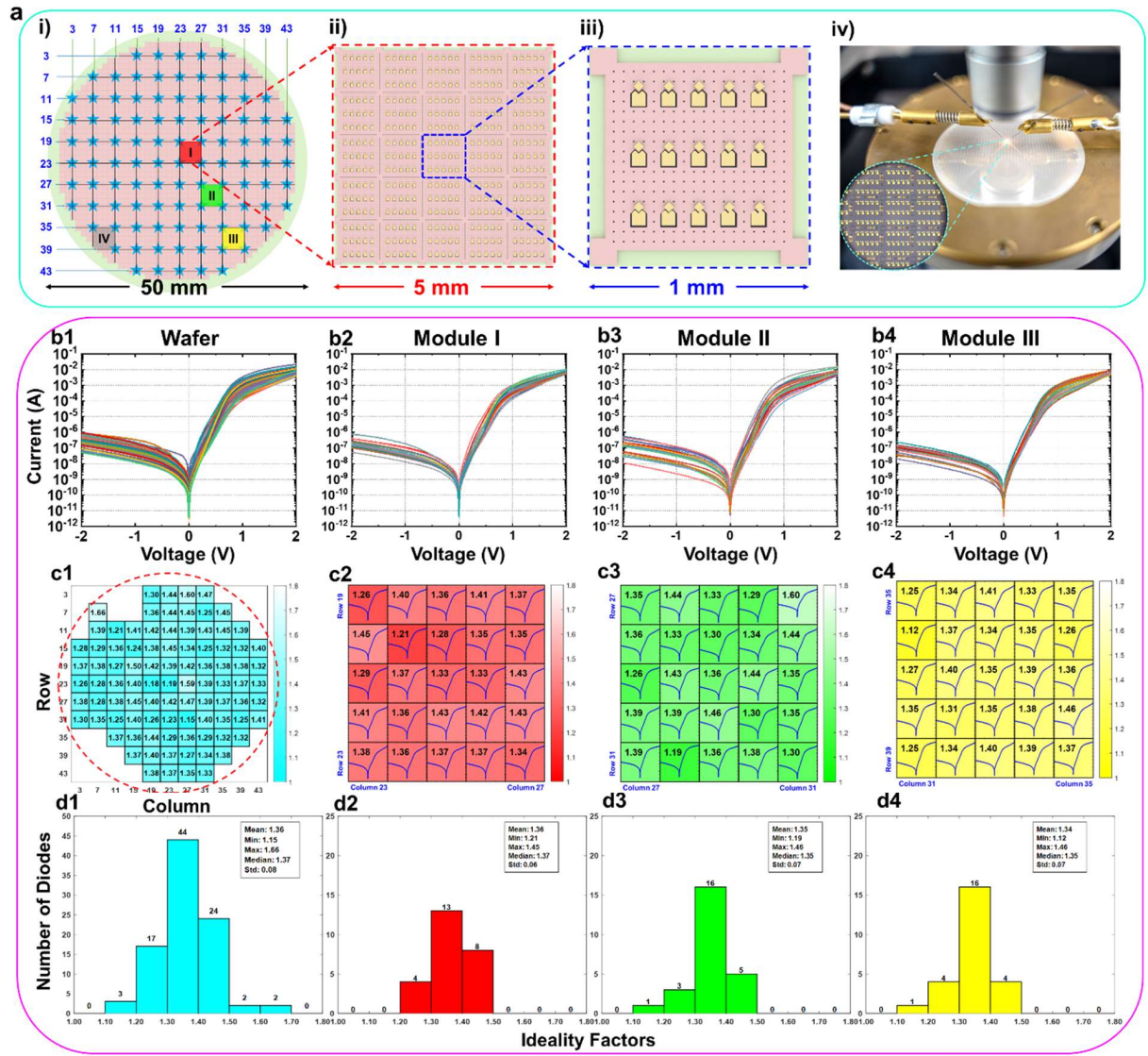

**Fig. 4 Schematic illustration, I-V characteristics, and statistical analysis of wafer-scale Si/GaN device mapping.** (a) Illustration of device distribution on the 2-inch GaN wafer. i) Top view of the distribution of the grafted Si NMs across the entire 2-inch GaN wafer, ii) the Si NMs are divided into multiple modules, with each module containing 5 × 5 sections, and iii) each section represents a single unit of Si NMs, with a size of 1 mm by 1 mm and containing 15 diodes. iv) Photograph of the wafer mounted on the probe station for the mapping measurement. (b)-(d) Summary of global and module mapping results for the wafer. (b) Compiled I-V curves for the b1) entire wafer, b2) module I, b3) module II, and b4) module III from left to right, respectively. (c) I.F. mapping schematics of the c1) entire wafer, c2) module I, c3) module II, and c4) module III from left to right, respectively. (d) Statistical results of I.F. values d1) across the

entire wafer and within d2) module I, d3) module II, and d4) module III, presented from left to right.

**Figure 4** shows the I-V characteristics of the fabricated grafted Si/GaN *p-n* diodes measured across the full wafer. Given the wafer-scale size and large number of diodes, we developed a statistically robust mapping strategy—a three-tiered "wafer–module–section" method—illustrated in **Fig. 4(a)**. The wafer contains 1,581 Si NM sections (**Fig. 4(a i)**). To ensure representative sampling, we selected sections marked with blue stars, covering the majority of the wafer. These Si NMs are divided into modules, each a 5 × 5 grid of sections (**Fig. 4(a ii)**), with each section a 1 mm × 1 mm Si NM unit containing fifteen Si/GaN diodes (**Fig. 4(a iii)**). **Figure 4(a iv)** shows the wafer mounted on a probe station for mapping.

**Figures 4(b1)** to **4(b4)** present the collected I-V curves, representing the entire wafer and modules I, II, and III, respectively. We achieved an overall high yield of functional devices (93.88%), with 92 out of 98 devices exhibiting consistent rectifying behavior. Device failure was primarily attributed to Si NM displacement, causing misalignment in the top left sections of the wafer, or incomplete Si NM transfer, leading to a loss of Si/GaN diodes in the bottom left sections, as shown in **Figs. 4(c1)** and **4(b3)**. Despite these minor issues, the I-V curves demonstrate remarkable consistency and uniformity across the entire wafer and within individual modules. See **Supplementary Information** for schematical representation of yield variation analysis during grafting and device fabrication phases.

**Figures 4(c1)** to **4(c4)** display the extracted ideality factor (I.F.) values derived from the corresponding I-V curves in **Figs 4(b1)** to **4(b4)**, with **Figs. 4(d1)** to **4(d4)** presenting the statistical analyses of these values. We observed consistently low I.F. values, with a mean value of 1.36 globally and mean values of 1.36, 1.35, and 1.34 for modules I, II, and III, respectively. The corresponding standard deviations—0.08 globally and 0.06, 0.07, and 0.07 for modules I, II, and III, respectively—confirm the excellent uniformity of the I.F. distribution, both globally and locally. Additionally, we mapped module IV, located in the bottom left corner, to demonstrate the distribution of both functional and short-circuited devices at the wafer's fringe (see **Supplementary Information** for details); although module IV shows a higher concentration of nonfunctional devices in the global mapping (**Fig. 4(c1)**), localized mapping within the module reveals a substantial number of functional devices. See **Supplementary Information** for further section-level mapping and $I_{ON}/I_{OFF}$ ratio statistics.

After demonstrating the successful fabrication and characterization of wafer-scale Si/GaN *p-n* diodes, the scalability of these devices was further investigated through capacitance-voltage (C-V) measurements, with the results summarized in **Fig. 5**. Device footprints were scaled into five sizes, ranging from 27,908 µm² to 256,333 µm², corresponding to the areas of the circular mesas (D1 to D5), as shown in the insets of **Figs. 5(a1)** to **5(a5)**. See **Supplementary Information** for tabulated mesa area parameters. **Figures 5(a1)** to **5(a5)** present the measured C-V and calculated 1/C²-V characteristics for each device footprint across different frequencies. These C-V curves show minimal frequency dispersion across all device sizes. As shown in **Fig. 5(b)**, device D1 underwent dense frequency sweeping, revealing a frequency-independent response from 10 kHz to 2 MHz under various biases, indicating well-suppressed interfacial density of states ($D_{it}$). A more comprehensive analysis and evaluation of the $D_{it}$ value from our experimental results can be found in **Supplementary Information. Figs. 5(c)** and **5(e)** show the combined C-V and 1/C²-V characteristics, respectively, for all five device sizes. The measured junction capacitance increases proportionally with the mesa (junction) area; after normalizing the capacitance (C) by the junction area (A) (**Figs. 5(d)** and **5(f)**), the unit capacitance (C/A) remains largely independent of device size, from D1 to D5 (27,908 µm² to 256,333 µm²). Additionally, the built-in potential of the grafted Si/GaN diodes, as extracted from the normalized 1/C²-V curves (**Fig. 5(f)**), consistently measures around 1.5 V across all five devices. The frequency dispersion-free C-V behavior across devices with footprints scaled by ~10X, along with the monotonic increase in capacitance, confirms the scalability of the devices and the uniformity of the heterojunction interface quality.

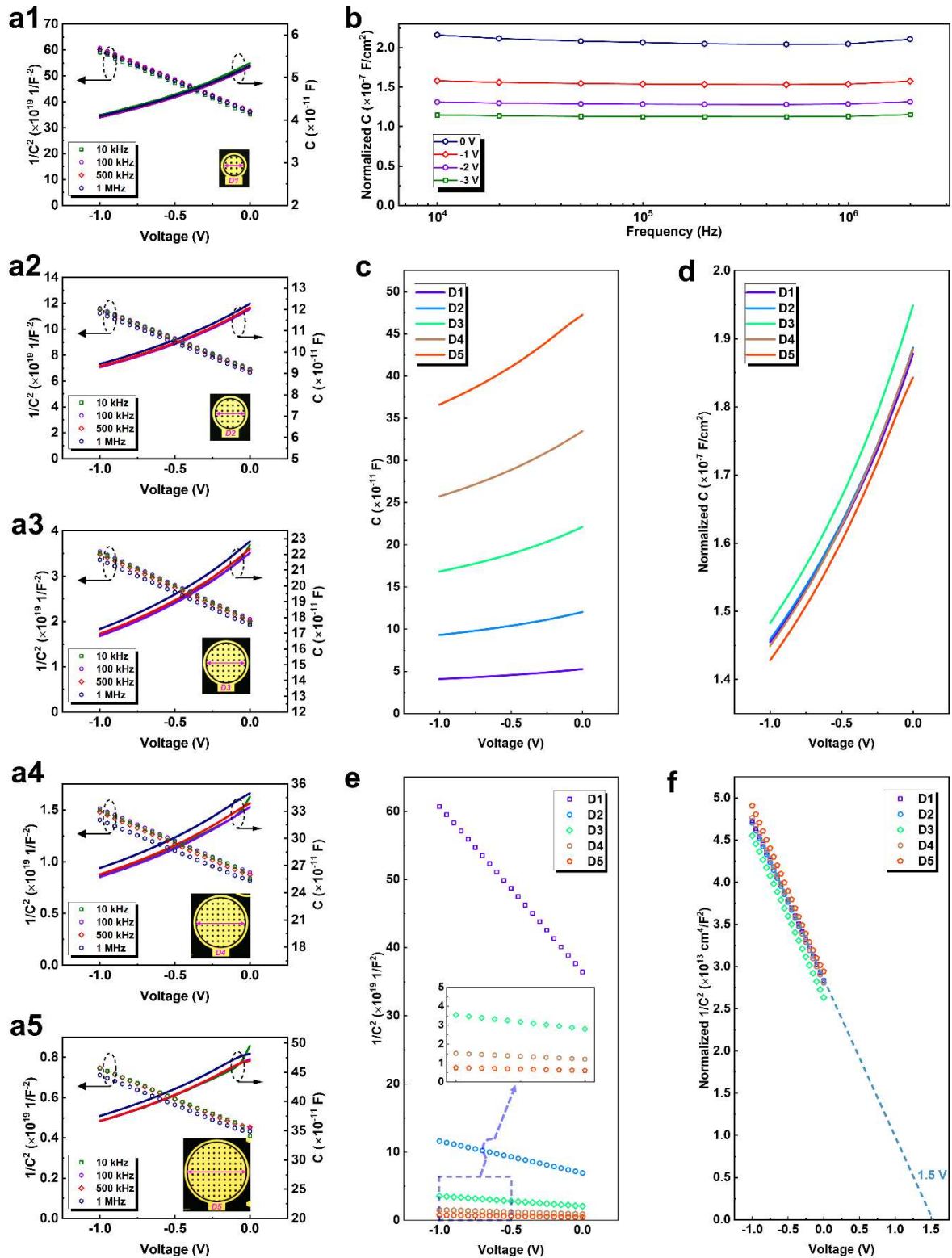

**Fig. 5 Capacitance-voltage (C-V) characterization of the fabricated Si/GaN heterojunction diodes.** (a1) to (a5) C-V characteristics of devices D1 to D5 with varying footprints, measured at different frequencies. (b) Frequency sweep of device D1, demonstrating that the measured junction capacitance remains dispersion-free from 10 kHz to 2 MHz. (c) Combined C-V curves for all five device sizes. (d) C-V curves normalized by device footprint, demonstrating area-independent behavior. (e) $1/C^2$-V curves for the five devices. (f) Normalized $1/C^2$-V curves for built-in voltage extraction, and further confirming the area-independent characteristic of the devices.

## Conclusion

This work establishes wafer-scale semiconductor grafting as a powerful technology for fabricating high-quality, single-crystalline, and electrically active heterojunctions, overcoming the limitations of conventional techniques such as wafer bonding and heteroepitaxy, which require strict lattice and/or thermal expansion matching. The wafer-scale grafting process demonstrated versatility, reliability, and repeatability, enabling successful integration of various material combinations, including Si/Si, Si/Sapphire, and Si/GaN. X-ray diffraction and scanning transmission electron microscopy confirmed the crystallinity of these heterojunctions, revealing atomically clean and sharp interfaces. Wafer-scale *p-n* diode fabrication and mapping further validated the scalability and uniformity of this approach, with capacitance-voltage measurements confirming size- and frequency-independent characteristics, indicative of a high-quality interface with minimal density of states. Unlike earlier millimeter-scale grafting methods, which often resulted in inconsistent yields and performance due to the dependence on localized quality of NMs, wafer-scale grafting significantly improves uniformity and production yield by enabling lift-off, transfer printing, and device fabrication across entire wafers, ensuring consistent quality and performance. This scalable approach opens new avenues for integrating diverse heterogeneous electronic and photonic devices—such as diodes, HBTs, LEDs, and lasers—onto a single wafer, facilitating the design of more efficient, customizable, and high-performance chips. Overall, this work provides a robust foundation for advancing semiconductor grafting technology toward industrial-scale applications and the widespread commercialization of heterogeneous semiconductor devices.

## Experiment Section

*Description of fabrication of wafer-scale Si/GaN heterojunction*

The process begins with patterning and dry etching of pixelized and anchored Si NMs (see **Supplementary Information** for detailed pixelization design), followed by a precisely controlled hydrofluoric acid (HF) undercut process to detach Si NMs from the Si handling substrate. This undercut process maintains anchor support, protecting the Si NMs from wrinkles, delamination, and displacement. Meanwhile, the destination GaN substrate is cleaned using standard procedures to remove particles, residual chemicals, and native oxides, and is subsequently passivated by ultrathin $Al_2O_3$ via an ALD process. After the undercut is finished, the freestanding Si NMs on the original Si substrate are removed from the HF solution, rinsed and dried. A honeycomb-textured PDMS stamp is then mounted onto an MJB-3 aligner (see **Supplementary Information** for detailed honeycomb structure design). After the PDMS is fully attached to the Si NMs, a prompt pick-up action releases them for transfer-printing to the destination substrate using the PDMS stamp. Finally, grafted Si/GaN heterostructures are created through a rapid thermal annealing (RTA) process, which firmly bonds the Si NMs to the ALD-passivated destination substrates by forming chemical bonds at the interface.

*Description of fabrication of Si/GaN heterojunction sample for XPS characterization*

To study the interface properties, *i.e.*, the band alignment of the Si/GaN heterojunction using XPS analysis, the thickness of the top Si NM should be reduced from the original 180 nm to around 10 nm without changing its crystallinity. This thickness reduction was achieved by a two-step process. First, the Si NM was etched down using an RIE etcher. After dry etching, the Si NM thickness was further fine-tuned using an in-situ ion bombardment through the ion gun equipped with the XPS tool. Finally, a completed 10 nm Si/GaN heterojunction sample is ready for XPS characterization. This XPS sample is named "grafted Si/GaN heterojunction".

*Description of fabrication of wafer-scale Si/GaN heterojunction p-n diodes*

In the fabrication process demonstration, a single Si NM pixel is used to represent the entire wafer, providing a clearer illustration of the wafer-scale *p-n* diode fabrication, as shown in **Fig. 6**. The process is divided into four stages. First, trenches are photolithographically patterned, followed by a two-step etching process to expose the $n^+$ GaN layer (**Fig. 6(d)**). Second, cathode metallization is performed on the exposed $n^+$ GaN, followed by rapid thermal annealing (RTA) to form ohmic contact (**Fig. 6(e)**). Third, anode metallization is carried out on the $p^+$ Si NM (**Fig. 6(f)**). Finally, interconnected devices are isolated through Si NM dry etching, with the anode metal acting as a hard mask, as shown in **Fig. 6(g1)**, with a magnified view of the completed device in **Fig. 6(g2)**.

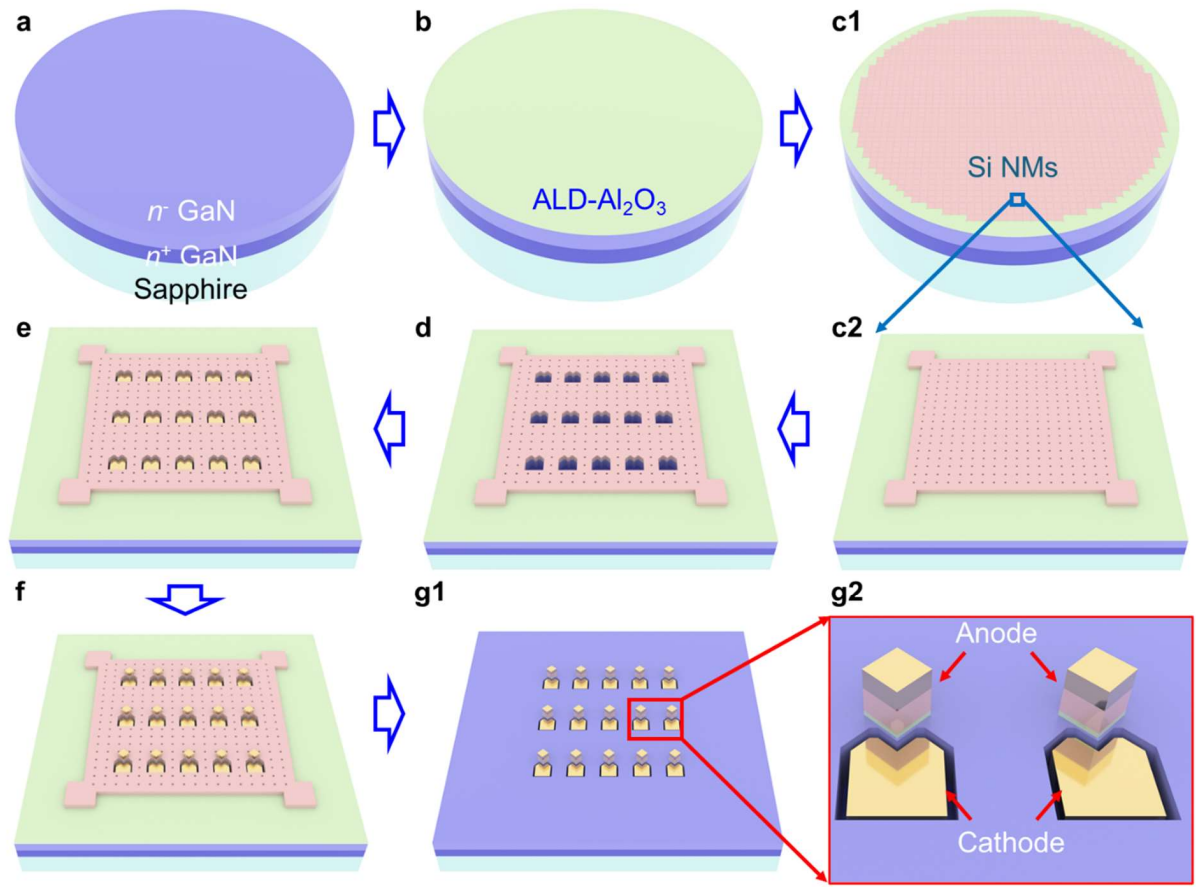

**Fig. 6. Process flow illustration of wafer-scale Si/GaN heterojunction *p-n* diodes.** (a) – (b) The as-grown GaN wafer is passivated with ALD after pre-cleaning. (c1) Wafer-scale grafting is performed to form the Si/GaN heterostructure, with a magnified view of a Si NM section shown in (c2). (d) A trench is first patterned by etching into the $n^+$ GaN layer, on which cathode metallization is performed, shown in (e), followed by an anode metallization on Si NMs (f). (g1) The anode metal is used as a shadow mask to isolate devices by dry etching the exposed extra Si NMs. A magnified view of the fabricated Si/GaN diode is shown in (g2).

*XPS characterization of Si/GaN heterojunction*

The XPS measurements were performed by a Thermo Scientific K Alpha X-ray Photoelectron Spectrometer (XPS) with an Al Kα X-ray source (h*ν*=1486.6 eV). Detailed parameters were exactly same as those from our previous work [9,25].

*XRD characterization of Si/GaN heterojunction*

The XRD 2theta-omega spectrum was measured by a Malvern Panalytical Empyrean X-ray diffractometer with a Cu K-α X-ray source.

*STEM characterization of Si/GaN heterojunction*

The cross-sectional TEM specimens were prepared using two methods. 1) A conventional TEM sample preparation method combining mechanical polishing procedure and a final ion milling step with PIPS 691 precision ion polishing system; and 2) a Focused ion beam (FIB) cutting and lift-out method using Thermo Scientific Helios G4 UX Dual Beam. Scanning transmission electron microscopy (STEM) micrographs taken under high-angle annular dark field (HAADF) mode and energy-dispersive X-ray spectroscopy (EDX) data were acquired by an aberration corrected microscope (ThermoFisher Themis Z Double Corrected 300 kV FEG S/TEM) and a Thermo Fisher Scientific TALOS F200X S/TEM system.

## Acknowledgements

The work was partially supported by DARPA H2 program under grant: HR0011-21-9-0109.